\documentclass[a4paper]{mem}
\usepackage{natbib}
\usepackage{psfig}
\usepackage{graphicx}
\usepackage[a4paper]{hyperref}
\idline{73}{23}
\begin{document}
   \title{The Cartwheel ring under X-ray light}

   \author{A. Wolter\inst{1} \and
   G. Trinchieri \inst{1}
}

   \offprints{A. Wolter}
\mail{anna@brera.mi.astro.it}

   \institute{\inst{1}INAF - Osservatorio Astronomico di Brera
via Brera 28, 20121 Milano 
             }

   \abstract{The Cartwheel, archetypical ring galaxy with strong star 
formation activity concentrated in a peculiar annular structure, has
been imaged with the CHANDRA ACIS-S instrument. We present here preliminary
results, that confirm the high luminosity detected earlier with the ROSAT HRI.
Many bright isolated sources are visible in the star-forming ring. 
A diffuse component at luminosity of the order of 10$^{40}$ cgs is 
also detected.
   }
   \authorrunning{A. Wolter and G. Trinchieri}
   \titlerunning{The X-ray Cartwheel}
   \maketitle
%

\section{The Cartwheel}
The Cartwheel (A0035-324) is a spectacular ring galaxy, located in a tight group
at z $\sim$ 0.03, of about 0.2 Mpc physical size. The ring-shaped galaxy 
formed after a head-on
collision with a compact galaxy (one of the 4 members of the
group): the gravitational rebound generates a transient density wave
that propagates outward and forms the ring. 
When/if the gas swept by the
density wave reaches a ``critical'' density in the ring a process of
intense star formation is triggered (eg. \cite{Hernquist}).

\begin{figure}
\label{char}
\hskip 0.3cm \psfig{file=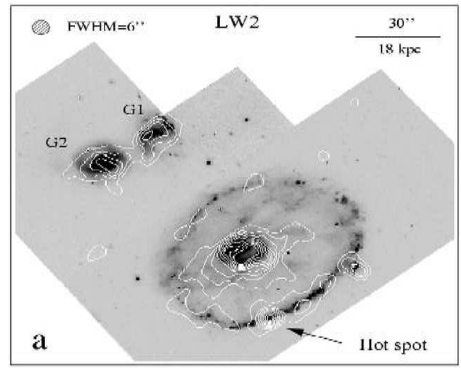,width=6.0truecm,height=4.8truecm}
\caption{The ISO contours at 7$\mu m$ superposed on the HST image of the 
Cartwheel and its two close companions G1 and G2 (adapted from
\cite{charmandaris}).}
\end{figure}

Optical observations have revealed the presence of exceptionally luminouos HII
regions, with an estimated supernova rate as high as 1 SN/yr (\cite{fosbury}).
A recent starburst in the outer ring of the Cartwheel is confirmed by
the evidences provided by the data at 
many wavelenghts, in particular at H$\alpha$ where many HII regions are
detected, and in the infrared, where ISO detects a hot spot in the
south portion of the ring (\cite{charmandaris}, see Figure~\ref{char}).

The first X-ray detection of the Cartwheel was obtained with the 
ROSAT HRI (\cite{wolter}): the total flux of
F$_x$=6.5 $\times 10^{-14}$ cgs corresponds to an
un-expectedly high luminosity
of L$_x$ = 5.0 $\times 10^{41}$ cgs, in the ROSAT band
(we use H$\rm_0$ = 50 km s$^{-1}$ Mpc$^{-1}$, for a distance of 180 Mpc and a
scale of 0.834 kpc/arcsec). We note that these numbers are virtually 
insensitive, in the ROSAT band, to the choice of spectral model that had
to be assumed a priori due to the lack of spectral resolution of the HRI.
The outer ring was clearly the sole site of emission, and the relatively
high luminosity was interpreted as due to the enhanced level of star
formation.

\begin{figure*}
\centerline{\psfig{file=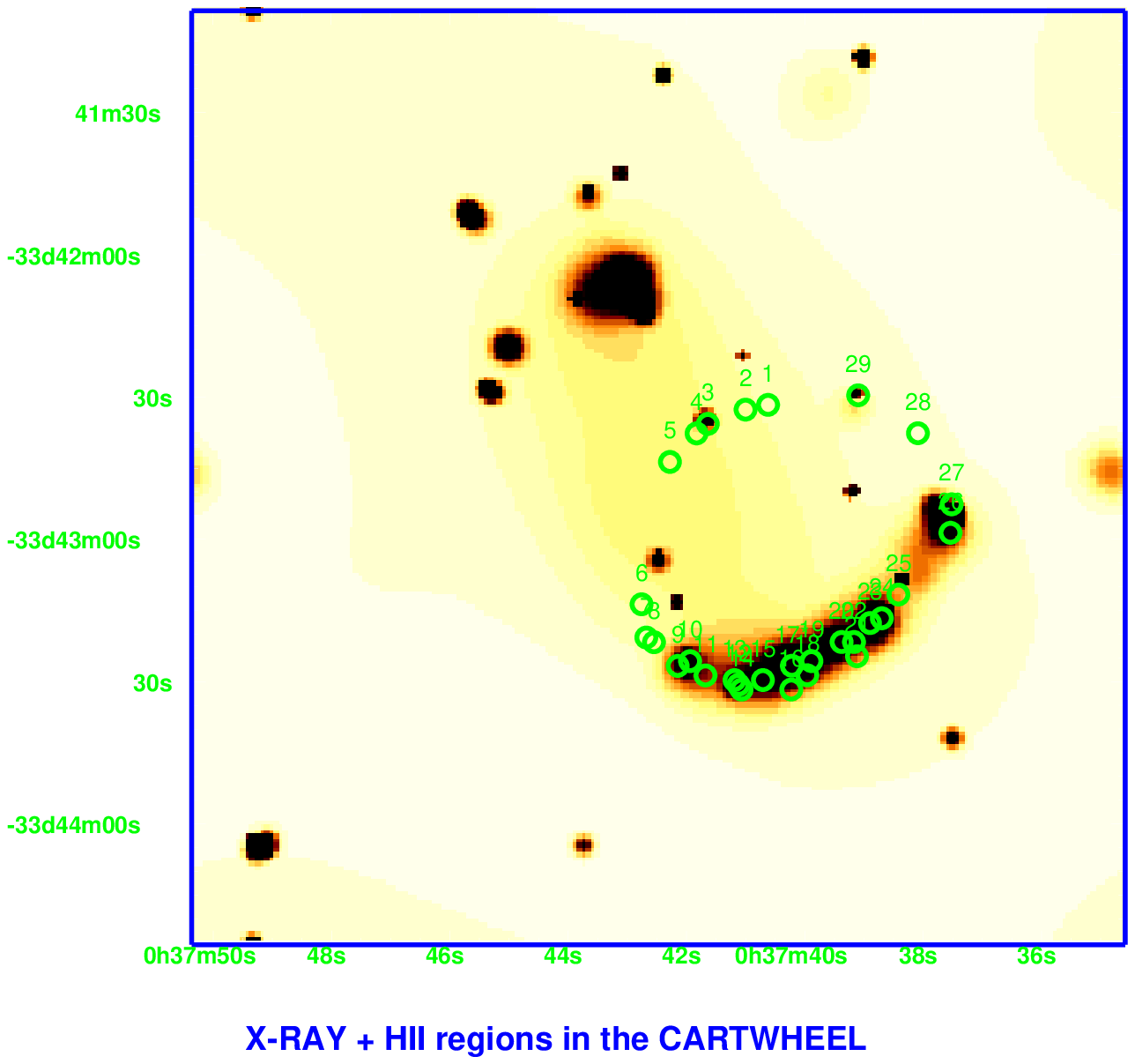,width=12.truecm}}
\caption{The ACIS-S image, smoothed with {\it csmooth}. The circles
mark the position of the HII knots measured by \cite{higdon}. }
\label{hii}
\end{figure*}

\section{Chandra Images}
We obtained in AO2 a CHANDRA ACIS-S 
77 ksec exposure. (For a description of the CHANDRA mission
see~\cite{weiss}).
The CHANDRA image (see Figure~\ref{hii}) shows that the X-ray emission 
is extended and clumpy, mostly associated with the outer ring.
The emission is stronger in the Southern quadrant, where the massive 
and luminous HII regions at large H$\alpha$ luminosities are found 
(\cite{higdon}).
The ROSAT detection of the two nearby galaxies G1 and G2 is 
confirmed.

The smoothed X-ray data from the ACIS-S CHANDRA detector (see 
Figure~\ref{hii}) show that the many individual sources in the ring
appear to be surrounded by a diffuse component.
There is a slight indication of diffuse emission also in the center of 
the galaxy, but with no direct connection to the inner ring or the 
spokes.  The nucleus is however clearly not detected, at a level
of 10$^{39}$ cgs. 
Giant HII regions and complex structures, typically coincident with peaks
of H$\alpha$ emission, have been observed in actively star-forming objects 
like the interacting system ``The Antennae'' (NGC 4038/9; 
eg.~\cite{fabbiano}) as extremely bright X-ray sources, with intrinsic
luminosities reaching several $\times 10^{40}$ cgs. To check this association
in the Carthweel, we have plotted the position of the HII knots as 
measured by \cite{higdon} onto the X-ray image in Figure~\ref{hii}. 
The positions of the circles that mark the HII regions have been
shifted by about 1" in RA and 0.5" in dec for better agreement with
the X-ray peaks (well within the positional uncertainty of both
the X-ray and the HII reference frame). A general trend in the locus 
of the X-ray and HII emission is evident. However, there is no one-to-one 
correspondence.
The brightest individual source in the outer ring has a luminosity
L$_X \sim 5 \times 10^{40}$ cgs in the ROSAT (0.2-2.4 keV) band, of
the same order of those found in the Antennae.

\section{Chandra spectra}
From inspection of images in two different energy cuts we found that a
diffuse emission is present, mostly in the soft band and most probably
not confined to the ring only. We therefore decided to investigate
the two components (isolated sources and diffuse emission) separately.
We extracted a combined spectrum from all the individual sources, that
we consider as point-like.  The statistics is not such that a study of
the individual profiles of the single sources will be significant,
but in total we have more than 2500 net counts for the sum of the sources. 
Spectral data are binned so that each bin has a significance $\geq 2 \sigma$.

\begin{figure}
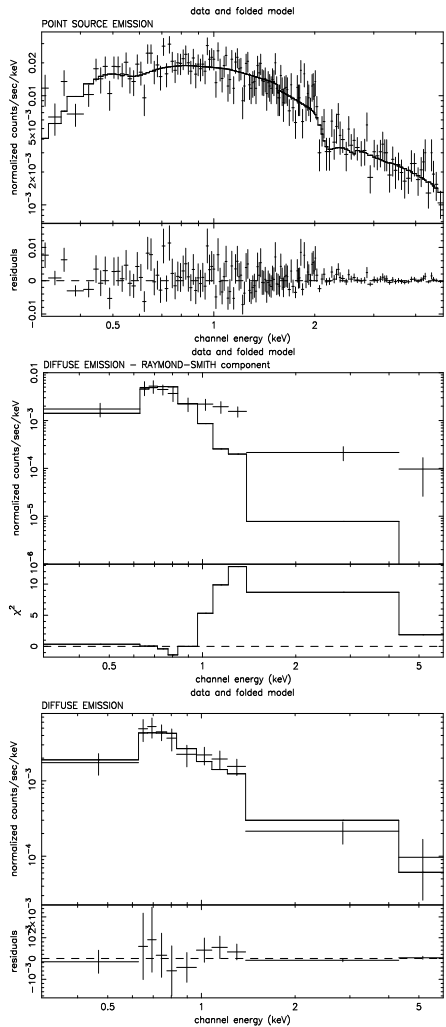

\label{sp}
\psfig{file=Wolter3a.ps,width=5.8truecm,height=4.5truecm,angle=-90}
\psfig{file=Wolter3b.ps,width=5.8truecm,height=4.5truecm,angle=-90}
\psfig{file=Wolter3c.ps,width=5.8truecm,height=4.5truecm,angle=-90}
\caption{{\it Top} ACIS spectrum of the combined individual unresolved
sources. The solid line corresponds to a power law fit with
$\Gamma=1.4$ and  N$_{\rm H}$ = 9.1 $\times 10^{20}$ 
cm$^{-2}$. {\it Center} ACIS spectrum of the diffuse component:
even if the statistics is low, the spectrum can not be fitted by a single
component. We show here the fit with a hot plasma component only: the
hard tail is evident. The same kind of residuals is obtained by the fit
with a power law component only.
{\it Bottom} The same ACIS spectrum for the diffuse component: we show
here the fit with a hot plasma component plus
power law (see text for details).}
\end{figure}

\begin{figure}
\psfig{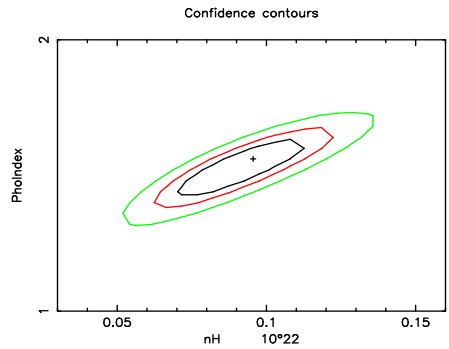}
\caption{Confidence contours of the two parameters $\Gamma$ and N$_{\rm H}$
for the point-source spectrum.}
\label{conf}
\end{figure}

The spectrum of the sum of the individual sources is shown in 
Figure~\ref{sp} ({\it Top}). The data are fitted by 
a power law with $\Gamma$ = 1.4 and N$_{\rm H}$ = 9.1 $\times 10^{20}$ 
cm$^{-2}$. The derived low energy absorption is higher than the galactic
one, but consistent with the reddening observed in the HII regions.
The contour plot of the uncertainties in spectral index and
low energy absorption is shown in Figure~\ref{conf}. The photon index we 
obtain is in the range observed in other bright binaries in nearby galaxies 
when fitted with a simple power law model (eg.~\cite{hump}).
The measured unabsorbed flux is 1.8$\times 10^{-13}$ cgs (0.5-50 keV band) that
corresponds to a Luminosity of 7.3$\times 10^{41}$ (0.5-5.0 keV) cgs. 

The spectrum shown in Figure~\ref{sp} ({\it Bottom}) for 
the ``extended''
emission is derived from a region that includes virtually all of the 
Cartwheel extent, but excludes the point sources in the ring. While with only
about 500 counts a successful detailed analysis is not guaranteed,
we found that the spectrum can only be fitted by a complex model (at
least two components, see Figure~\ref{sp} {\it Center}). 
The best fit is a combination of a
power law ($\Gamma$=1.6, N$_{\rm H}$=1.3$\times 10^{21}$ cm$^{-2}$ plus 
a plasma model (Raymond-Smith component with solar --fixed-- abundance 
and kT=0.3 keV). 

The flux of the power law component, representing non-resolved
binaries, is F$_x$ = 1.3$\times 10^{-14}$ cgs (0.5-5.0 keV), about
10\% of the resolved point source flux.
The flux of the diffuse hot gas component is F$_x$ = 4. $\times 10^{-15}$
cgs (0.5-50 keV). 
It contributes mostly at 0.6-0.9 keV as expected from the temperature
found.  The total Luminosity of the gas is of the order of 
$1-2 \times 10^{40}$ cgs.

\section{Conclusions}

We have presented preliminary results from a CHANDRA observation of the
Cartwheel. A number of isolated sources is present, coincident with the
region of high star formation detected at other wavelengths. A diffuse 
gaseous component might permeate the entire system, but the detection is
not on firm statistical grounds. The isolated sources have large X-ray
luminosities, reaching a total of at least L$_X = 7.3 \times 10^{41}$ cgs 
in the 0.5-5.0 keV band. Future follow-up XMM data might help in deriving the
spectra of the brightest individual sources thus confirming their identity,
and firmly establish the presence and extent of the diffuse emission.
Detailed analysis of the CHANDRA data will be presented in a forthcoming 
paper (Wolter et al. in preparation.)

\begin{acknowledgements}

This work has received partial financial support from the Italian 
Space Agency.
\end{acknowledgements}

\bibliographystyle{aa}

\end{document}